\documentstyle[12pt,twoside]{article}

\def\greaterthansquiggle{\raise.3ex\hbox{$>$\kern-.75em\lower1ex\hbox{$\sim$}}}
\def\lessthansquiggle{\raise.3ex\hbox{$<$\kern-.75em\lower1ex\hbox{$\sim$}}}
\newcommand{\beq}{\begin{equation}}
\newcommand{\eeq}{\end{equation}}
\newcommand{\beqa}{\begin{eqnarray}}
\newcommand{\eeqa}{\end{eqnarray}}
\newcommand{\beqan}{\begin{eqnarray*}}
\newcommand{\eeqan}{\end{eqnarray*}}
\newcommand{\ba}{\begin{array}}
\newcommand{\ea}{\end{array}}

\def\nz{\ifmmode {I\hskip -3pt N} \else {\hbox {$I\hskip -3pt N$}}\fi}
\def\zz{\ifmmode {Z\hskip -4.8pt Z} \else
       {\hbox {$Z\hskip -4.8pt Z$}}\fi}
\def\qz{\ifmmode {Q\hskip -5.0pt\vrule height6.0pt depth 0pt
       \hskip 6pt} \else {\hbox
       {$Q\hskip -5.0pt\vrule height6.0pt depth 0pt\hskip 6pt$}}\fi}
\def\rz{\ifmmode {I\hskip -3pt R} \else {\hbox {$I\hskip -3pt R$}}\fi}
\def\cz{\ifmmode {C\hskip -4.8pt\vrule height5.8pt\hskip 6.3pt} \else
       {\hbox {$C\hskip -4.8pt\vrule height5.8pt\hskip 6.3pt$}}\fi}

\newtheorem{theorem}{Theorem}
\newtheorem{definition}{Definition}

\normalsize
\begin{document}

\begin{titlepage}
\begin{center}
{\Large \bf 
 ON QUANTUM CAPACITY AND ITS BOUND
}\\[24pt]
Masanori Ohya$^*$ and Igor V. Volovich$^{**}$\\
$^*$ Department of Information Sciences \\
Tokyo University of Science \\
Noda City, Chiba 278-8510,
Japan
\\
$^{**}$
Steklov Mathematical Institute,\\
Gubkin St.8, GSP-1 117966, \\
Moscow, Russia
\vfill
{\bf Abstract} \\
\end{center}
The quantum capacity of a pure quantum channel and that of
classical-quantum-classical channel are discussed in detail based on the
fully quantum mechanical mutual entropy. It is proved that the quantum
capacity generalizes the so-called Holevo bound.
\vfill
\small
e-mail: ohya@is.noda.tus.ac.jp
\end{titlepage}
\section{Introduction}

Measure theoretic formulation of the mutual entropy (information) in
classical systems was done by Kolmogorov \cite{Kol} and Gelfand, Yaglom \cite%
{GY}, which enabled to define the capacity of information channel. In
quantum systems, there have been several definitions of the mutual entropy
for classical input and quantum output \cite{Hol,Ing,Lev}. In 1983, Ohya
defined \cite{O2} the fully quantum mechanical mutual entropy, i.e., for
quantum input and quantum output, by means of the relative entropy of
Umegaki \cite{Ume}, and he extended it \cite{O4} to general quantum systems
by using the relative entropy of Araki \cite{Ara} and Uhlmann \cite{Uhl}. In
this short note, we prove that the quantum capacity \cite{O5} of a quantum
channel derived from the fully quantum mechanical mutual entropy generalizes
the so-called Holevo bound.

\section{\textbf{Mutual Entropy}}

The quantum mutual entropy was introduced in \cite{O2} for a quantum input
and quantum output, namely, for a purely quantum channel, and it was
generalized for a general quantum system described by C*-algebraic
terminology \cite{O4}. We briefly review the mutual entropy in usual quantum
system described by a Hilbert space.

Let $\mathcal{H}$ be a Hilbert space for an input space,\textbf{\ }$B%
\mathcal{(H)}$ be the set of all bounded linear operators on $\mathcal{H}$
and $\mathcal{\ S(H)}${\scriptsize \ }be the set of all density operators on 
$\mathcal{H}.$ An output space is described by another Hilbert space $%
\tilde{\mathcal{H}}$ , but often $\mathcal{H=}
\widetilde{\mathcal{H}}$. A channel from the input system to the output
 system is a
mapping $\Lambda $* from $\mathcal{S(H)}$ to 
$\mathcal{S(\widetilde{H})}$ \cite{O1}. A channel $\Lambda $* is said to be 
completely
positive if the dual map $\Lambda $ satisfies the following condition: $%
\Sigma _{k,j=1}^{n}$ $A_{k}^{\ast }\Lambda (B_{k}^{\ast }B_{j})A_{j}\geq 0$
for any $n\in $\textbf{N} and any $A_{j}\in B({\mathcal H}), 
B_{k} \in B(\widetilde{{\mathcal H}})$.

An input state $\rho $ $\in \mathcal{S(H)}$ is sent to the output system
through a channel $\Lambda $*, so that the output state is written as $%
\tilde{\rho }\equiv \Lambda ^{\ast }\rho .$ Then it is important to
ask how much information of $\rho $ is sent to the output state $\Lambda
^{\ast }\rho .$ This amount of information transmitted from input to output
is expressed by the quantum mutual entropy.

The quantum mutual entropy was introduced on the basis of the von Neumann
entropy $\left( S(\rho )\equiv -tr\rho \log \rho \right) $ for purely
quantum communication processes. The mutual entropy depends on an input
state $\rho $ and a channel $\Lambda ^{\ast }$, so it is denoted by $I\left(
\rho ;\Lambda ^{\ast }\right) $, which should satisfy the following
conditions:

(i) The quantum mutual entropy is well-matched to the von Neumann entropy.
Furthermore, if a channel is trivial, i.e., $\Lambda ^{*}=$ identity map,
then the mutual entropy equals to the von Neumann entropy: $I\left( \rho
;id\right) $ = $S\left( \rho \right) $.

(ii) When the system is classical, the quantum mutual entropy reduces to
classical one.

(iii) Shannon's fundamental inequality \cite{Sha} 0$\leq $ $I\left( \rho
;\Lambda ^{*}\right) \leq S\left( \rho \right) $ is held.

To define such a quantum mutual entropy extending Shannon's and
Gelefand-Yaglom's classical mutual entropy, we need the quantum relative
entropy and the joint state (it is called ''compound state'' in the sequel)
describing the correlation between an input state $\rho $ and the output
state $\Lambda ^{\ast }\rho $ through a channel $\Lambda ^{\ast }$. A finite
partition of measurable space in classical case corresponds to an orthogonal
decomposition $\left\{ E_{k}\right\} $ of the identity operator I of $%
\mathcal{H}$ in quantum case because the set of all orthogonal projections
is considered to make an event system in a quantum system. It is known \cite%
{OP} that the following equality holds

\begin{eqnarray*}
\mbox{sup} \left\{ -\sum_{k}tr\rho E_{k}\log tr\rho E_{k};\left\{ E_{k}\right\}
\right\} =-tr\rho \log \rho ,
\end{eqnarray*}
and the supremum is attained when $\left\{ E_{k}\right\} $ is a Schatten
decomposition\cite{Sch} of $\rho .$ 
Therefore the Schatten decomposition is used to
define the compound state and the quantum mutual entropy following the
formulation of the classical mutual entropy by Kolmogorov, Gelfand and
Yaglom \cite{GY}.

The compound state $\sigma _{E}$ (corresponding to joint state in CS) of $%
\rho $ and $\Lambda ^{*}\rho $ was introduced in \cite{O2,O3}, which is
given by

\begin{equation}
\sigma _{E}=\sum_{k}\lambda _{k}E_{k}\otimes \Lambda ^{\ast }E_{k},
\end{equation}
where $E$ stands for a Schatten decomposition $\left\{ E_{k}\right\} $ of $%
\rho ,$ so that the compound state depends on how we decompose the state $%
\rho $ into basic states (elementary events), in other words, how to observe
the input state.

The relative entropy for two states $\rho $ and $\sigma $ is defined by
Umegaki \cite{Ume} and Lindblad \cite{Lin}, which is written as

\begin{equation}
S\left( \rho ,\sigma \right) =\left\{ 
\begin{array}{ll}
tr\rho \left( \log \rho -\log \sigma \right) & \left( \mbox{when }\overline{%
ran\rho }\subset \overline{ran\sigma }\right) \\ 
\infty & \left( \mbox{otherwise}\right)%
\end{array}
\right.
\end{equation}

Then we can define the mutual entropy by means of the compound state and the
relative entropy \cite{O2}, that is,

\begin{equation}
I\left( \rho ;\Lambda ^{\ast }\right) =\sup \left\{ S\left( \sigma _{E},\rho
\otimes \Lambda ^{\ast }\rho \right) ;E=\left\{ E_{k}\right\} \right\} ,
\label{defofQME}
\end{equation}
where the supremum is taken over all Schatten decompositions because this
decomposition is not unique unless every eigenvalue is not degenerated. Some
computations reduce it to the following form \cite{O2}:

\begin{equation}
I\left( \rho ;\Lambda ^{\ast }\right) =\sup \left\{ \sum_{k}\lambda
_{k}S\left( \Lambda ^{\ast }E_{k},\Lambda ^{\ast }\rho \right) ;E=\left\{
E_{k}\right\} \right\} .  \label{QME2}
\end{equation}
This mutual entropy satisfies all conditions (i)$\sim $(iii) mentioned above.

It is important to note here that the Schatten decomposition of $\rho $ is
unique when the input system is classical. That is, when an input state $%
\rho $ is given by a probability distribution or a probability measure. For
the case of probability distribution ; $\rho =\left\{ \lambda _{k}\right\} ,$
the Schatten decomposition is uniquely given by

\begin{equation}
\rho =\sum_{k}\lambda _{k}\delta _{k},  \label{ClassicalSchatten}
\end{equation}
where $\delta _{k}$ is the delta measure;

\begin{equation}
\delta _{k}\left( j\right) =\delta _{k,j}=\{_{0(k\neq j)}^{1(k=j)},\forall j.
\end{equation}
Therefore for any channel $\Lambda ^{*},$ the mutual entropy becomes

\begin{equation}
I\left( \rho ;\Lambda ^{*}\right) =\sum_{k}\lambda _{k}S\left( \Lambda
^{*}\delta _{k},\Lambda ^{*}\rho \right) ,  \label{CME}
\end{equation}
which equals to the following usual expression of Shannon when the minus is
well-defined:

\begin{equation}
I\left( \rho ;\Lambda ^{\ast }\right) =S\left( \Lambda ^{\ast }\rho \right)
-\sum_{k}\lambda _{k}S\left( \Lambda ^{\ast }\delta _{k}\right) .
\end{equation}
The above equality has been taken as the definition of the mutual entropy
for a classical-quantum channel \cite{Hol,Ing,Lev}.

\smallskip Note that the definition (\ref{defofQME}) of the mutual entropy
is written as

\begin{eqnarray*}
I( \rho ;\Lambda ^{\ast }) =\sup \left\{ \sum_{k}\lambda
_{k}S\left( \Lambda ^{\ast }\rho _{k},\Lambda ^{\ast }\rho \right) ;\rho
=\sum_{k}\lambda _{k}\rho _{k}\in F_{o}\left( \rho \right) \right\} ,
\end{eqnarray*}
where $F_{o}\left( \rho \right) $ is the set of all orthogonal finite
decompositions of $\rho .$ The proof of the above equality is given in \cite%
{Mur-O} by means of fundamental properties of the quantum relative entropy.

\section{\textbf{Communication Processes}}

\textbf{\ }We discuss communication processes in this section\cite{Ing,OP}.
Let $A=\{a_{1,}a_{2,}\cdot \cdot ,a_{n}\}$ be a set of certain alphabets and 
$\Omega $ be the infinite direct product of $A:$ $\Omega =A^{Z}\equiv \Pi
_{-\infty }^{\infty }A$ calling a message space. In order to send a
information written by an element of this message space to a receiver, we
often need to transfer the message into a proper form for a communication
channel. This change of a message is called a coding. Precisely, a coding is
a measurable one to one map $\xi $ from $\Omega $ to a proper space $X$ .

Let $( \Omega ,{\mathcal F}_{\Omega},P(\Omega)) $ be an input
probability space and $X$ be the coded input space. This space $X$ may be a
classical object or a quantum object. For instance, $X$ is a Hilbert space $%
\mathcal{H}$ of a quantum system, then the coded input system is described
by $\left( B(\mathcal{H)},\mathcal{S}(\mathcal{H)}\right) $.

An output system is similarly described as the input system: The coded
output space is denoted by $\tilde{X}$ and the decoded output space
is $\tilde{\Omega }$ made by another alphabets. An transmission
(map) from $X$ to $\tilde{X}$ is described by a channel reflecting
all properties of a physical device, which is denoted by $\gamma $ here.
With a decoding $\tilde{\xi },$ the whole information transmission
process is written as 
\begin{equation}
\Omega \mapsto{\xi }X\mapsto{\gamma }
\tilde{X}\mapsto{\tilde{\xi }}
\widetilde{\Omega }.  \label{transmissionprocess}
\end{equation}
That is, a message $\omega \in \Omega $ is coded to $\xi \left( \omega
\right) $ and it is sent to the output system through a channel $\gamma $,
then the output coded message becomes $\gamma \circ \xi \left( \omega
\right) $ and it is decoded to $\tilde{\xi }\circ \gamma \circ \xi
\left( \omega \right) $ at a receiver.

This transmission process is mathematically set as follows: M messages are
sent to a receiver and the $k$th message $\omega ^{\left( k\right) }$ occurs
with the probability $\lambda _{k}.$ Then the occurrence probability of each
message in the sequence $\left( \omega ^{\left( 1\right) },\omega ^{\left(
2\right) },\cdot \cdot \cdot ,\omega ^{\left( M\right) }\right) $of M
messages is denoted by $\rho =\left\{ \lambda _{k}\right\} ,$ which is a
state in a classical system. If $\xi $ is a classical coding, then $\xi
\left( \omega \right) $ is a classical object such as an electric pulse. If $%
\xi $ is a quantum coding, then $\xi \left( \omega \right) $ is a quantum
object (state) such as a coherent state. Here we consider such a quantum
coding, so that $\xi \left( \omega ^{\left( k\right) }\right) $ is a quantum
state, and we denote $\xi \left( \omega ^{\left( k\right) }\right) $ by $%
\sigma _{k}.$ Thus the coded state for the sequence $\left( \omega ^{\left(
1\right) },\omega ^{\left( 2\right) },\cdot \cdot \cdot ,\omega ^{\left(
M\right) }\right) $ is written as

\begin{equation}
\sigma =\sum_{k}\lambda _{k}\sigma _{k}.
\end{equation}
This state is transmitted through a channel $\gamma .$ This channel is
expressed by a completely positive mapping $\Gamma ^{*},$ in the sense of
Sec.1, from the state space of $X$ to that of $\tilde{X}$ , hence
the output coded quantum state $\tilde{\sigma }$ is $\Gamma
^{*}\sigma .$ Since the information transmission process can be understood
as a process of state (probability) change, when $\Omega $ and $
\widetilde{\Omega }$ are classical and $X$ and $\tilde{X}$ are quantum,
the process (\ref{transmissionprocess}) is written as

\begin{equation}
P\left( \Omega \right) \mapsto{\Xi ^{\ast }}{\mathcal S}%
\left( {\mathcal H}\right) \mapsto{\Gamma^{\ast }}%
{\mathcal S(\widetilde{H})}\mapsto{\widetilde{\Xi }%
^{\ast }} P( \widetilde{\Omega } ),
\label{qtransmission}
\end{equation}
where $\Xi ^{\ast }$ $($resp.$\widetilde{\Xi }^{\ast })$ is the channel
corresponding to the coding $\xi $ (resp.$\widetilde{\xi }$ ) and $%
{\mathcal S}\left( {\mathcal H}\right) $ (resp.${\mathcal S(\widetilde{%
H}}))$ is the set of all density operators (states) on $\mathcal{%
\ H}$ (resp.$\widetilde{\mathcal{H}}$ $)$.

We have to be care to study the objects in the above transmission process (%
\ref{transmissionprocess}) or (\ref{qtransmission}). Namely, we have to make
clear which object is going to study. For instance, if we want to know the
information capacity of a quantum channel $\gamma (=\Gamma ^{*}),$ then we
have to take $X$ so as to describe a quantum system like a Hilbert space and
we need to start the study from a quantum state in quantum space $X\ $not
from a classical state associated to a message. If we like to know the
capacity of the whole process including a coding and a decoding, which means
the capacity of a channel $\tilde{\xi }\circ \gamma \circ \xi (=%
\tilde{\Xi }^{*}\circ \ \Gamma ^{*}\circ \Xi ^{*})$, then we have
to start from a classical state$.$ In any case, when we concern the capacity
of channel, we have only to take the supremum of the mutual entropy $I\left(
\rho ;\Lambda ^{*}\right) $ over a quantum or classical state $\rho $ in a
proper set determined by what we like to study with a channel $\Lambda ^{*}.$
We explain this more precisely in the next section.

\section{\textbf{Channel Capacity }}

We discuss two types of channel capacity in communication processes, namely,
the capacity of a quantum channel $\Gamma ^{*}$ and that of a classical
(classical-quantum-classical) channel $\tilde{\Xi }^{*}\circ \
\Gamma ^{*}\circ \Xi ^{*}.$

(1) \textit{Capacity of quantum channel:} The capacity of a quantum channel
is the ability of information transmission of a quantum channel itself, so
that it does not depend on how to code a message being treated as a
classical object and we have to start from an arbitrary quantum state and
find the supremum of the mutual entropy. One often makes a mistake in this
point. For example, one starts from the coding of a message and compute the
supremum of the mutual entropy and he says that the supremum is the capacity
of a quantum channel, which is not correct. Even when his coding is a
quantum coding and he sends the coded message to a receiver through a
quantum channel, if he starts from a classical state, then his capacity is
not the capacity of the quantum channel itself. In his case, usual Shannon's
theory is applied because he can easily compute the conditional distribution
by usual (classical) way. His supremum is the capacity of a
classical-quantum-classical channel, and it is in the second category
discussed below.

Let ${\mathcal S}_{0}(\subset $ $\mathcal{S(H))}$ be the set of all states
prepared for expression of information. Then the capacity of the channel $%
\Gamma ^{\ast }$ with respect to ${\mathcal S}_{0}$ is defined as:

\begin{definition}
The capacity of a quantum channel $\Gamma ^{\ast }$ is 
\begin{equation}
C^{{\mathcal S}_{0}}\left( \Gamma ^{\ast }\right) =\sup \{I\left( \rho
;\Gamma ^{\ast }\right) ;\rho \in {\mathcal S}_{0}\}.  \label{defofcapacity}
\end{equation}
Here $I\left( \rho ;\Gamma ^{\ast }\right) $ is the mutual entropy given in (%
\ref{defofQME}) or (\ref{QME2}) with $\Lambda ^{\ast }=\Gamma ^{\ast }.$
\end{definition}

When ${\mathcal S}_{0}=\mathcal{S(H)}$ , $C^{\mathcal{S}(\mathcal{H)}}\left(
\Gamma ^{\ast }\right) $ is denoted by $C\left( \Gamma ^{\ast }\right) $ for
simplicity. In \cite{O4,OPW1,Mur-O}, we also considered the pseudo-quantum
capacity $C_{p}\left( \Gamma ^{\ast }\right) $ defined by (\ref%
{defofcapacity}) with the pseudo-mutual entropy $I_{p}\left( \rho ;\Gamma
^{\ast }\right) $ where the supremum is taken over all finite decompositions
instead of all orthogonal pure decompositions: 
\begin{equation}
I_{p}\left( \rho ;\Gamma ^{\ast }\right) =\sup \left\{ \sum_{k}
\lambda_{k}S\left( \Gamma ^{\ast }\rho _{k},\Gamma ^{\ast }\rho \right) ;\rho
=\sum_{k}\lambda _{k}\rho _{k},\mbox{ finite decomposition}\right\} .
\end{equation}
However the pseudo-mutual entropy is not well-matched to the conditions
explained in Sec.2, and it is difficult to compute numerically \cite{OPW2}.
From the monotonicity of the mutual entropy \cite{OP}, we have

\begin{eqnarray*}
0\leq C^{{\mathcal S}_{0}}\left( \Gamma ^{*}\right) \leq C_{p}^{
{\mathcal S}_{0}}\left( \Gamma ^{*}\right) \leq \sup \left\{ S(\rho );\rho \in 
{\mathcal S}_{0}\right\} .
\end{eqnarray*}

(2) \textit{Capacity of classical-quantum-classical channel:} The capacity
of C-Q-C channel $\tilde{\Xi }^{*}\circ \ \Gamma ^{*}\circ \Xi ^{*} 
$ is the capacity of the information transmission process starting from the
coding of messages, therefore it can be considered as the capacity including
a coding (and a decoding). As is discussed in Sec.3, an input state $\rho $
is the probability distribution $\left\{ \lambda _{k}\right\} $ of messages,
and its Schatten decomposition is unique as (\ref{ClassicalSchatten}), so
the mutual entropy is written by (\ref{CME}):

\begin{equation}
I\left( \rho ;\tilde{\Xi }^{\ast }\circ \ \Gamma ^{\ast }\circ \Xi
^{\ast }\right) =\sum_{k}\lambda _{k}S\left( \tilde{\Xi }^{\ast
}\circ \ \Gamma ^{\ast }\circ \Xi ^{\ast }\delta _{k},\tilde{\Xi }%
^{\ast }\circ \ \Gamma ^{\ast }\circ \Xi ^{\ast }\rho \right) .
\end{equation}
If the coding $\Xi ^{\ast }$ is a quantum coding, then $\Xi ^{\ast }\delta
_{k}$ is expressed by a quantum state. Let denote the coded quantum state by 
$\sigma _{k}$ and put $\sigma =\Xi ^{\ast }\rho =\sum_{k}\lambda _{k}\sigma
_{k}.$ We denote the set of such quantum codings by $\mathcal{C}$. Then the
above mutual entropy becomes

\begin{equation}
I\left( \rho ;\tilde{\Xi }^{\ast }\circ \ \Gamma ^{\ast }\circ \Xi
^{\ast }\right) =\sum_{k}\lambda _{k}S\left( \tilde{\Xi }^{\ast
}\circ \ \Gamma ^{\ast }\sigma _{k},\tilde{\Xi }^{\ast }\circ \
\Gamma ^{\ast }\sigma \right) .  \label{CQCME}
\end{equation}
This is the expression of the mutual entropy of the whole information
transmission process starting from a coding of classical messages. Hence the
capacity of C-Q-C channel is as follows:

\begin{definition}
The capacity of C-Q-C channel is 
\begin{equation}
C^{P_{0}}\left( \tilde{\Xi }^{\ast }\circ \ \Gamma ^{\ast }\circ
\Xi ^{\ast }\right) =\sup \{I\left( \rho ;\tilde{\Xi }^{\ast }\circ
\ \Gamma ^{\ast }\circ \Xi ^{\ast }\right) ;\rho \in P_{0}\},
\end{equation}
where $P_{0}(\subset P(\Omega ))$ is the set of all probability
distributions prepared for input (a-priori) states (distributions or
probability measures).
\end{definition}

Moreover the capacity for coding free in $\mathcal{C}$ is found by taking
the supremum of the mutual entropy (\ref{CQCME}) over all probability
distributions in $P_{0}$ and all codings in $\mathcal{C}$:

\begin{equation}
C_{c}^{P_{0}}\left( \tilde{\Xi }^{\ast }\circ \ \Gamma ^{\ast
}\right) =\sup \{I\left( \rho ;\tilde{\Xi }^{\ast }\circ \ \Gamma
^{\ast }\circ \Xi ^{\ast }\right) ;\rho \in P_{0},\Xi ^{\ast }\in \mathcal{C}%
\}.
\end{equation}
There are several ways to decode quantum states such as quantum
measurements, so that denote such decodings by $\mathcal{D}$. The capacity
for decoding free in $\mathcal{D}$ is

\begin{equation}
C_{d}^{P_{0}}\left( \ \Gamma ^{\ast }\circ \Xi ^{\ast }\right) =\sup
\{I\left( \rho ;\tilde{\Xi }^{\ast }\circ \ \Gamma ^{\ast }\circ
\Xi ^{\ast }\right) ;\rho \in P_{0},\widetilde{\Xi }^{\ast }\in 
{\mathcal D}\}.
\end{equation}
The last capacity is for both coding and decoding free and it is given by

\begin{equation}
C_{cd}^{P_{0}}\left( \ \Gamma ^{\ast }\right) =\sup \{I\left( \rho ;
\widetilde{\Xi }^{\ast }\circ \ \Gamma ^{\ast }\circ \Xi ^{\ast }\right) ;\rho
\in P_{0},\Xi ^{\ast }\in {\mathcal C},\widetilde{\Xi }^{\ast }\in 
\mathcal{D}\}.
\end{equation}
These capacities $C_{c}^{P_{0}},C_{d}^{P_{0}},$ $C_{cd}^{P_{0}}$ do not
measure the ability of the quantum channel $\Gamma ^{\ast }$ itself, but
measure the ability of $\Gamma ^{\ast }$ through the coding and decoding.
The above three capacities $C^{P_{0}},$ $C_{c}^{P_{0}},$ $C_{cd}^{P_{0}}$
satisfy the following inequalities 
\begin{eqnarray*}
0\leq C^{P_{0}}\left( \tilde{\Xi }^{\ast }\circ \ \Gamma ^{\ast
}\circ \Xi ^{\ast }\right) \leq C_{c}^{P_{0}}\left( \tilde{\Xi }%
^{\ast }\circ \ \Gamma ^{\ast }\right) ,\mbox{ }C_{d}^{P_{0}}\left( \ \Gamma
^{\ast }\circ \Xi ^{\ast }\right) \leq C_{cd}^{P_{0}}\left( \ \Gamma ^{\ast
}\right) \leq \sup \left\{ S(\rho );\rho \in P_{0}\right\}
\end{eqnarray*}
where $S(\rho )$ is not the von Neumann entropy but the Shannon entropy: -$%
\sum \lambda _{k}\log \lambda _{k}.$

Remark that if $\sum_{k}\lambda _{k}S(\Gamma ^{\ast }\sigma _{k})$ is
finite, then (\ref{CQCME}) becomes

\begin{equation}
I\left( \rho ;\tilde{\Xi }^{\ast }\circ \ \Gamma ^{\ast }\circ \Xi
^{\ast }\right) =S(\widetilde{\Xi }^{\ast }\circ \Gamma ^{\ast }\sigma
)-\sum_{k}\lambda _{k}S(\widetilde{\Xi }^{\ast }\circ \Gamma ^{\ast
}\sigma _{k}).
\end{equation}
Further, if $\rho $ is a probability measure having a density function $%
f(\lambda )$ and each $\lambda $ corresponds to a quantum coded state $%
\sigma (\lambda ),$ then $\sigma =\int f(\lambda )$ $\sigma (\lambda
)d\lambda $ and

\begin{equation}
I\left( \rho ;\tilde{\Xi }^{\ast }\circ \ \Gamma ^{\ast }\circ \Xi
^{\ast }\right) =S(\tilde{\Xi }^{\ast }\circ \Gamma ^{\ast }\sigma
)-\int f(\lambda )S(\tilde{\Xi }^{\ast }\circ \Gamma ^{\ast }\sigma
(\lambda ))d\lambda ,
\end{equation}
which is less than

\begin{equation}
S(\Gamma ^{\ast }\sigma )-\int f(\lambda )S(\Gamma ^{\ast }\sigma (\lambda
))d\lambda .
\end{equation}
This bound is computed in several cases\cite{OPW1,YO}. This bound is a
special one of the following inequality

\begin{eqnarray*}
I( \rho ;\tilde{\Xi }^{\ast }\circ \ \Gamma ^{\ast }\circ \Xi
^{\ast }) \leq I\left( \rho ;\ \Gamma ^{\ast }\circ \Xi ^{\ast
}\right) ,
\end{eqnarray*}
which comes from the monotonicity of the relative entropy. When the decoding
is not taken into account, then we have only to consider the mutual entropy $%
I\left( \rho ;\ \Gamma ^{\ast }\circ \Xi ^{\ast }\right) $ above.

Let us define an extension of the functional of the relative entropy. If $A$
and $B$ are two positive Hermitian operators (not necassarily the states,
i.e., not necessarily with unit traces) then we set 
\begin{eqnarray*}
S(A,B)=trA( \log A-\log B)
\end{eqnarray*}
There is the following \textit{Bogoliubov inequality}. 
\begin{eqnarray*}
S(A,B)\geq trA\left( \log trA-\log trB\right)
\end{eqnarray*}

The following theorem gives us the bound of the mutual entropy $I\left( \rho
;\ \Gamma ^{\ast }\circ \Xi ^{\ast }\right) $.

\begin{theorem}
For a probability distribution $\rho =\left\{ \lambda _{k}\right\} $ and a
quantum coded state $\sigma =\Xi ^{\ast }\rho \equiv \sum_{k}\lambda
_{k}\sigma _{k}$ , $\lambda _{k}\geq 0$, $\sum_{k}\lambda _{k}=1$, one has
the following inequality for any quantum channel decomposed as $\Gamma
^{\ast }=\Gamma _{1}^{\ast }\circ \Gamma _{2}^{\ast }$ such that $\Gamma
_{1}^{\ast }\sigma \equiv \sum_{i}E_{i}\sigma E_{i}$ by a projection valued
measure $\left\{ E_{i}\right\} :$ 
\begin{eqnarray}
\sum_{k}\lambda _{k}S(\sigma _{k},\sigma ) &\geq &I\left( \rho ;\ \Gamma
^{\ast }\circ \Xi ^{\ast }\right) =\sum_{k}\lambda _{k}S(\Gamma ^{\ast
}\sigma _{k},\Gamma ^{\ast }\sigma ) \\
&\geq &\sum_{i}\left[ -tr(\Gamma _{2}^{\ast }\sigma E_{i})\log tr(\Gamma
_{2}^{\ast }\sigma E_{i})+\sum_{k}\lambda _{k}tr(\Gamma _{2}^{\ast }\sigma
_{k}E_{i})\log tr(\Gamma _{2}^{\ast }\sigma _{k}E_{i})\right]  \nonumber
\end{eqnarray}
\end{theorem}

{\bf proof:}\\
The equality $I\left( \rho ;\ \Gamma ^{\ast }\circ \Xi ^{\ast }\right)
=\sum_{k}\lambda _{k}S(\Gamma ^{\ast }\sigma _{k},\Gamma ^{\ast }\sigma )$
is the case of the equality (15), and the first inequality comes from the
monotonicity of the relative entropy. Further by applying again the
monotonicity of the relative entropy, we have 
\begin{eqnarray*}
\sum_{k}\lambda _{k}S(\Gamma ^{\ast }\sigma _{k},\Gamma ^{\ast }\sigma )\geq
\sum_{k}\sum_{i}\lambda _{k}S(E_{i}\Gamma _{2}^{\ast }\sigma
_{k}E_{i},E_{i}\Gamma _{2}^{\ast }\sigma _{k}E_{i})
\end{eqnarray*}
\begin{eqnarray*}
\geq \sum_{k,i}\lambda _{k}tr(E_{i}\Gamma _{2}^{\ast }\sigma
_{k}E_{i})\left( \log tr(E_{i}\Gamma _{2}^{\ast }\sigma _{k}E_{i})-\log
tr(E_{i}\Gamma _{2}^{\ast }\sigma E_{i})\right)
\end{eqnarray*}
\begin{eqnarray*}
=\sum_{i}\left( -tr(\Gamma _{2}^{\ast }\sigma E_{i})\log tr(\Gamma
_{2}^{\ast }\sigma E_{i})+\sum_{k}\lambda _{k}tr(\Gamma _{2}^{\ast }\sigma
_{k}E_{i})\log tr(\Gamma _{2}^{\ast }\sigma _{k}E_{i})\right)
\end{eqnarray*}
Here the second inequality is due to the Bogoliubov inequality.
{\it Q.E.D.}
\par
In the case that the channel $\Gamma _{2}^{\ast }$ is trivial; $\Gamma
_{2}^{\ast }\sigma =\sigma ,$ the above inequality reduces to the bound
obtained by Holevo \cite{Hol}: 
\begin{eqnarray*}
\sum_{k}\lambda _{k}S(\sigma _{k},\sigma ) &=&-tr\sigma \log \sigma
+\sum_{k}\lambda _{k}tr\sigma _{k}\log \sigma _{k} \\
&\geq &I\left( \rho ;\ \Gamma ^{\ast }\circ \Xi ^{\ast }\right) =I\left(
\rho ;\ \Gamma _{1}^{\ast }\circ \Xi ^{\ast }\right) \\
&\geq &\sum_{i}\left[ -tr(\sigma E_{i})\log tr(\sigma
E_{i})+\sum_{k}\lambda _{k}tr(\sigma _{k}E_{i})\log tr(\sigma _{k}E_{i})%
\right]
\end{eqnarray*}
Remark that the right hand side in the inequality is sometimes called the 
\textit{accessible information}.

Using the above upper and lower bounds of the mutual entropy, we can compute
these bounds of the capacity in many different cases.
\\

\end{document}